\documentclass[pra,aps,showpacs,twocolumn]{revtex4}

\usepackage{amsmath}
\usepackage{bm}
\usepackage{graphicx}

\begin{document}

\title{Connection between rotation and miscibility in a two-component
Bose-Einstein condensate}

\author{Takayuki Shimodaira$^1$}
\author{Tetsuo Kishimoto$^2$}
\author{Hiroki Saito$^1$}

\affiliation{
$^1$Department of Engineering Science,
University of Electro-Communications, Tokyo 182-8585, Japan \\
$^2$Center for Frontier Science and Engineering, University of
Electro-Communications, Tokyo 182-8585, Japan
}

\date{\today}

\begin{abstract}
A two-component Bose-Einstein condensate rotating in a toroidal trap is
investigated.
The topological constraint depends on the density distribution of each
component along the circumference of the torus, and therefore the
quantization condition on the circulation can be controlled by changing
the miscibility using the Feshbach resonance.
We find that the system exhibits a variety of dynamics depending on the
initial angular momentum when the miscibility is changed.
\end{abstract}


\maketitle

\section{Introduction}

A Bose-Einstein condensate (BEC) in a toroidal
trap~\cite{Gupta,Arnold,Ryu,Olson} is an ideal system to study
fascinating properties of a superfluid, such as persistent
flow~\cite{Ryu}, symmetry breaking
localization~\cite{Carr,Kanamoto1,Kavo1,Parola}, and various rotating
states arising from quantized circulations~\cite{Kanamoto2,Kavo2}.

The quantization of circulation in superfluids originates from the
single-valuedness of the wave function, i.e., the change in the phase of
the wave function must be an integer multiple of $2\pi$ along a closed
path~\cite{Onsager,Feynman}.
An angular momentum of a BEC in a toroidal trap is therefore quantized
if the density is uniform.
However, if a density vanishes at a part of the circumference of the
torus, the phase can jump at the density defect~\cite{Carr,Drummond},
and the system is allowed to rotate with an arbitrary circulation.

\begin{figure}[t]
\includegraphics[width=7cm]{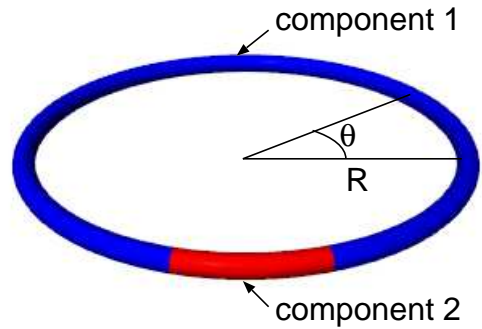}
\caption{
(color online) Schematic illustration of a phase-separated two-component
condensate in a toroidal trap.
}
\label{f:schematic}
\end{figure}

In the present paper, we consider a system of a two-component BEC
rotating in a toroidal trap~\cite{Smy,Malet}.
Let us consider a situation in which the repulsive interaction separates
the two components along the circumference of the torus as illustrated
in Fig.~\ref{f:schematic}.
Since the closed path along the torus for one component is blocked by
the other component, the quantization condition is not imposed on the
circulation around the torus, and the system can rotate at an arbitrary
frequency.
If we decrease the intercomponent repulsion from this initial state, the
two components mix, and the density distribution of each component
becomes multiply connected, and consequently, the quantization condition
is imposed on the circulation along the torus for each component.
During this dynamics, the total angular momentum must be conserved.
Nontrivial dynamics is thus expected due to the interplay between the
change of the quantization condition on the circulation and the angular
momentum conservation.

This paper is organized as follows.
Section~\ref{s:1D} studies one-dimensional (1D) ring geometry.
Section~\ref{s:formulate} gives formulation of the problem.
Section~\ref{s:1Dground} numerically investigates the stationary states
as a function of the intercomponent repulsion for a given angular
momentum, and Sec.~\ref{s:1Dbogo} calculates the Bogoliubov spectrum.
Section~\ref{s:1Danalytic} compares analytic results with numerical
ones.
Section~\ref{s:3D} demonstrates the dynamics of a two-component Rb BEC
in a 3D toroidal trap.
Section~\ref{s:conc} gives conclusions to this study.

\section{One dimensional ring}
\label{s:1D}

\subsection{Formulation of the problem}
\label{s:formulate}

In the mean-field approximation, a two-component BEC in a frame rotating
at a frequency $\Omega$ around the $z$ axis is described by the
Gross-Pitaevskii (GP) equations,
\begin{subequations} \label{GP}
\begin{eqnarray}
i \hbar \frac{\partial \psi_1}{\partial t} & = & \left(
-\frac{\hbar^2 \nabla^2}{2m_1} - \Omega L_z + V_1 + g_{11} |\psi_1|^2
+ g_{12} |\psi_2|^2 \right) \psi_1, \nonumber \\
\\
i \hbar \frac{\partial \psi_2}{\partial t} & = & \left(
-\frac{\hbar^2 \nabla^2}{2m_2} - \Omega L_z + V_2 + g_{22} |\psi_2|^2 +
g_{12} |\psi_1|^2 \right) \psi_2, \nonumber \\
\end{eqnarray}
\end{subequations}
where $\psi_j$ is the macroscopic wave function for the $j$th component
$(j = 1, 2)$ normalized as $\int |\psi_j|^2 d\bm{r} = N_j$ with $N_j$
being the number of atoms, $m_j$ is the atomic mass, $V_j$ is the
potential, and $L_z = -i \hbar \partial / \partial \theta$ is the $z$
component of the angular momentum operator with $\theta = {\rm arg} (x +
i y)$.
The interaction parameters are given by
\begin{equation}
g_{jj'} = 2 \pi \hbar^2 a_{jj'} \left( m_j^{-1} + m_{j'}^{-1} \right),
\end{equation}
where $a_{jj'}$ is the $s$-wave scattering length between atoms in
components $j$ and $j'$.
The condition for the phase separation in an infinite system is given by
$g_{11} g_{22} < g_{12}^2$ for $g_{jj'} > 0$~\cite{Pethick}.

In this section, for simplicity, we reduce the problem to 1D with a
periodic boundary condition.
We also assume $m_1 = m_2 \equiv m$, $V_1 = V_2$, and $g_{11} =
g_{22}$.
Assuming that the system is tightly confined in the torus and neglecting
excitations in the radial and $z$ directions, we obtain the normalized
GP equation as $(j \neq j')$
\begin{equation} \label{nGP}
i \frac{\partial \tilde{\psi}_j}{\partial \tau} = -\frac{1}{2}
 \frac{\partial^2 \tilde{\psi}_j}{\partial \theta^2} + i \tilde\Omega 
\frac{\partial \tilde{\psi}_j}{\partial \theta} + \tilde g_{jj}
|\tilde\psi_j|^2 \tilde \psi_j + \tilde g_{jj'} |\tilde\psi_{j'}|^2
\tilde \psi_j,
\end{equation}
where the wave function $\tilde \psi_j(\theta)$ is normalized as $\int
|\tilde \psi_j|^2 d\theta = N_j / (N_1 + N_2) \equiv n_j$, $\tau =
\hbar t / (m R^2)$, $\tilde\Omega = m R^2 \Omega / \hbar$, and $\tilde
g_{jj'} = m R^2 g_{jj'} \rho_{\rm av} / \hbar^2$ with $R$ being the
radius of the torus and $\rho_{\rm av}$ the averaged density.

\subsection{Stationary state with fixed angular momentum}
\label{s:1Dground}

We consider a situation in which a phase separated stationary state in
a rotating frame is prepared and then the intercomponent interaction
$g_{12} > 0$ is adiabatically decreased, during which the angular
momentum of the system is conserved.
We therefore seek the stationary state of Eq.~(\ref{nGP}) for a given
angular momentum as a function of $g_{12}$.

We use the imaginary time propagation method, in which $i$ on the
left-hand side of Eq.~(\ref{nGP}) is replaced by $-1$.
We add $-\tilde\mu_j |\tilde \psi_j|^2$ to the right-hand side of
Eq.~(\ref{nGP}) and control $\tilde\mu_j$ and $\tilde \Omega$ in the
imaginary time propagation such that the norm $\int |\tilde \psi_j|^2
d\theta = n_j$ in each component and the total angular momentum,
\begin{equation} \label{L}
\tilde L = \tilde L_1 + \tilde L_2 = -i \sum_{j = 1,2} \int \tilde
 \psi_j^* \frac{\partial \tilde \psi_j}{\partial \theta} d\theta,
\end{equation}
are kept constant.
The wave function thus converges to the stationary state with given
norms $n_j$ and an angular momentum $\tilde L$.

\begin{figure}[t]
\includegraphics[width=8.5cm]{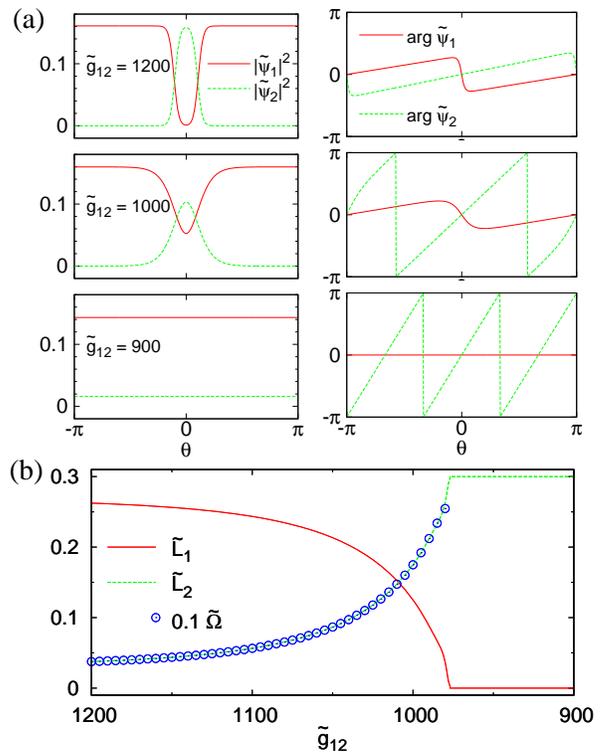}
\caption{
(color online) 
(a) Density distributions (left panels) and phase profiles (right
 panels) of the stationary states of Eq.~(\ref{nGP}) for $\tilde g_{12}
 = 1200$, $1000$, and $900$.
(b) Normalized angular momenta $\tilde L_1$ (solid
 curve), $\tilde L_2$ (dashed curve), and rotation frequency
 $\tilde\Omega$ (circles) of the stationary states as functions of
 $\tilde g_{12}$.
In (a) and (b), $\tilde g_{11} = \tilde g_{22} = 1000$ and the total
 angular momentum is fixed to $\tilde L = 0.3$.
The populations are $n_1 = 0.9$ and $n_2 = 0.1$.
}
\label{f:1D1}
\end{figure}
Figure~\ref{f:1D1} (a) shows the density and phase profiles of the
stationary states of Eq.~(\ref{nGP}), where the populations are $n_1 =
0.9$ and $n_2 = 0.1$ and the total angular momentum is fixed to $\tilde
L = 0.3$.
In Fig.~\ref{f:1D1}, we start from $\tilde g_{12} = 1200$ and
slowly decrease $\tilde g_{12}$ in the imaginary time propagation.
At $\tilde g_{12} = 1200$ [top panels of Fig.~\ref{f:1D1} (a)], the two
components are phase separated.
Component 2 breaks component 1 at $\theta = 0$, where the phase of
component 1 changes rapidly.
As $\tilde g_{12}$ is decreased, the two components are mixed, and the
density distribution of each component becomes uniform for $\tilde
g_{12} \lesssim 980$ [bottom panels of Fig.~\ref{f:1D1} (a)].
We note that the circulation,
\begin{equation}
\Gamma_j = \frac{1}{2\pi} \int_{-\pi}^\pi \frac{\partial {\rm arg}
 \tilde \psi_j}{\partial \theta} d\theta,
\end{equation}
changes due to the constraint of $\tilde L = 0.3$ being fixed.
As $\tilde g_{12}$ is decreased, component 1 goes to the uniform state
with $\Gamma_1 = 0$, and in order to maintain $\tilde L = 0.3$ the
circulation of component 2 increases to $\Gamma_2 = 3$, which satisfies
$n_1 \Gamma_1 + n_2 \Gamma_2 = \tilde L$.
Thus $\tilde L_2$ increases with a decrease in $\tilde L_1$ as shown by
the solid and dashed curves in Fig.~\ref{f:1D1} (b).
The plots in Fig.~\ref{f:1D1} (b) show the rotation frequency $\tilde
\Omega$ of the frame in which the density distributions are stationary.
We find that $\tilde\Omega$ coincides with $\tilde L_2 / n_2$,
indicating the rigid-body rotation of component 2 for $\tilde g_{12}
\lesssim 980$.

\begin{figure}[t]
\includegraphics[width=8.5cm]{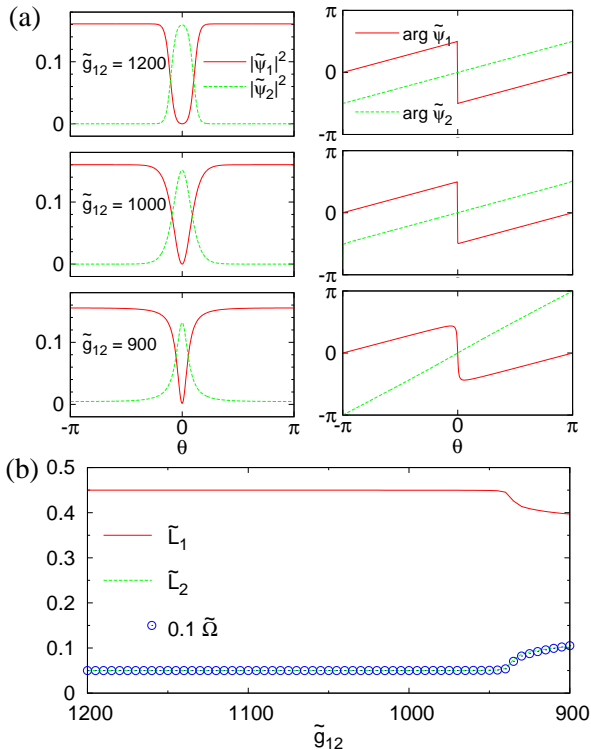}
\caption{
(color online) 
(a) Density distributions (left panels) and phase profiles (right
 panels) of the stationary states of Eq.~(\ref{nGP}).
(b) Normalized angular momenta $\tilde L_1$ (solid
 curve), $\tilde L_2$ (dashed curve), and rotation frequency
 $\tilde\Omega$ (circles) of the stationary states as functions of
 $\tilde g_{12}$.
The total angular momentum is fixed to $\tilde L = 0.5$.
The other conditions are the same as those in Fig.~\ref{f:1D1}.
}
\label{f:1D2}
\end{figure}
Figure~\ref{f:1D2} shows the case of $\tilde L = 0.5$.
Unlike in Fig.~\ref{f:1D1}, the dip in $|\tilde\psi_1|^2$ persists until
$\tilde g_{12} = 900$, which is occupied by the localized component 2.
The stability of this structure is due to the fact that ${\rm arg}
\tilde\psi_1$ jumps by $\pi$ at $\theta = 0$ and the stable dark soliton
is formed in component 1.
As $\tilde g_{12}$ is decreased, the tail of $|\tilde\psi_2|^2$ spreads
and the quantization condition is imposed on the circulation [bottom
panels of Fig.~\ref{f:1D2} (a)], which is the reason for the change in
$\tilde L_j$ at $\tilde g_{12} \simeq 950$ in Fig.~\ref{f:1D2} (b).

\begin{figure}[t]
\includegraphics[width=8.5cm]{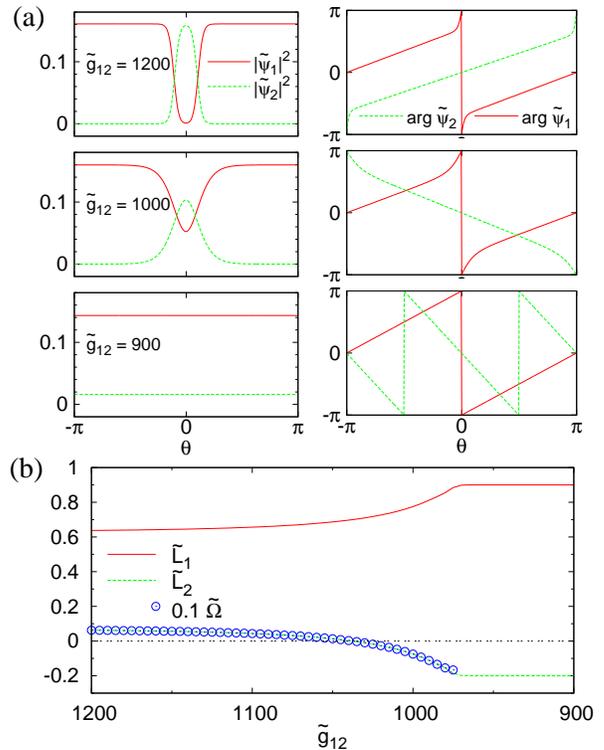}
\caption{
(color online) 
(a) Density distributions (left panels) and phase profiles (right
 panels) of the stationary states of Eq.~(\ref{nGP}).
(b) Normalized angular momenta $\tilde L_1$ (solid
 curve), $\tilde L_2$ (dashed curve), and rotation frequency
 $\tilde\Omega$ (circles) of the stationary states as functions of
 $\tilde g_{12}$.
The total angular momentum is fixed to $\tilde L = 0.7$.
The other conditions are the same as those in Fig.~\ref{f:1D1}.
}
\label{f:1D3}
\end{figure}
Figure~\ref{f:1D3} shows the case of $\tilde L = 0.7$.
As $\tilde g_{12}$ is decreased, $\tilde\psi_1$ goes to a uniform state
with circulation $\Gamma_1 = 1$ [bottom panels of Fig.~\ref{f:1D3} (a)],
which has an angular momentum $\tilde L_1 = n_1 = 0.9$.
As a consequence, $\tilde L_2$ must be $\tilde L - \tilde L_1 = -0.2$
and therefore $\Gamma_2$ becomes $-0.2 / n_2 = -2$.
Because of $\tilde L_2 \simeq n_2 \tilde \Omega$, the rotation frequency
of the system decreases with $\tilde g_{12}$.
This indicates that the rotation of the system slows down, stops, and
counterrotates in real time propagation as $\tilde g_{12}$ is decreased
adiabatically.
This interesting behavior is due to the interplay between the
quantization of circulation and the angular momentum conservation.

We note that the density distributions in Fig.~\ref{f:1D1} are exactly
the same as those in Fig.~\ref{f:1D3}.
This is because the GP equation (\ref{nGP}) is invariant under the
transformation,
\begin{subequations} \label{trans}
\begin{eqnarray}
\tilde\psi_j & \rightarrow & e^{i \ell \theta} \tilde\psi_j^*, \\
\tilde\Omega & \rightarrow & \ell - \tilde\Omega,
\end{eqnarray}
\end{subequations}
where $\ell$ is an integer.
By this transformation, the angular momentum is changed as
\begin{equation} \label{Ltrans}
\tilde L_j \rightarrow \ell n_j - \tilde L_j.
\end{equation}
The results in Fig.~\ref{f:1D3} agree with those obtained from
Fig.~\ref{f:1D1} by the transformation (\ref{trans}) and (\ref{Ltrans})
with $\ell = 1$.

\subsection{Bogoliubov analysis}
\label{s:1Dbogo}

If the time scale of the change in $g_{12}$ is much longer than the
inverse of the lowest excitation frequency, the wave function follows
the stationary states shown in Figs.~\ref{f:1D1}-\ref{f:1D3}.
In order to find the adiabatic condition, we perform the Bogoliubov
analysis.

We separate the wave function as
\begin{equation} \label{sep}
\tilde \psi_j(\theta, \tau) = [ \Psi_j(\theta) + \delta\psi_j(\theta,
\tau) ] e^{-i \tilde\mu_j \tau},
\end{equation}
where $\Psi_j$ is a stationary state of Eq.~(\ref{nGP}) and $\phi_j$ is
a small deviation.
Substituting Eq.~(\ref{sep}) with
\begin{equation}
\delta\psi_j(\theta, \tau) = u_j(\theta) e^{-i \tilde\omega \tau} +
 v_j^*(\theta) e^{i \tilde\omega \tau}
\end{equation}
into Eq.~(\ref{nGP}) and neglecting the second and higher orders of
$u_j$ and $v_j$, we obtain the Bogoliubov-de Gennes equation $(j \neq
j')$,
\begin{subequations} \label{BdG}
\begin{eqnarray}
& & -\frac{1}{2} \frac{\partial^2 u_j}{\partial\theta^2} + i \tilde\Omega
 \frac{\partial u_j}{\partial\theta} + \tilde g_{jj} (2 |\Psi_j|^2 u_j +
\Psi_j^2 v_j) \nonumber \\
& & + \tilde g_{jj'} (|\Psi_{j'}|^2 u_j + \Psi_j \Psi_{j'}^*
u_{j'} + \Psi_j \Psi_{j'} v_{j'}) - \mu_j u_j = \tilde\omega u_j,
\nonumber \\
\\
& & -\frac{1}{2} \frac{\partial^2 v_j}{\partial\theta^2} - i \tilde\Omega
 \frac{\partial v_j}{\partial\theta} + \tilde g_{jj} (2 |\Psi_j|^2 v_j +
\Psi_j^{*2} u_j) \nonumber \\
& & + \tilde g_{jj'} (|\Psi_{j'}|^2 v_j + \Psi_j^* \Psi_{j'}
v_{j'} + \Psi_j^* \Psi_{j'}^* u_{j'}) - \mu_j v_j = -\tilde\omega v_j.
\nonumber \\
\end{eqnarray}
\end{subequations}
We numerically diagonalize Eq.~(\ref{BdG}).

\begin{figure}[t]
\includegraphics[width=8.5cm]{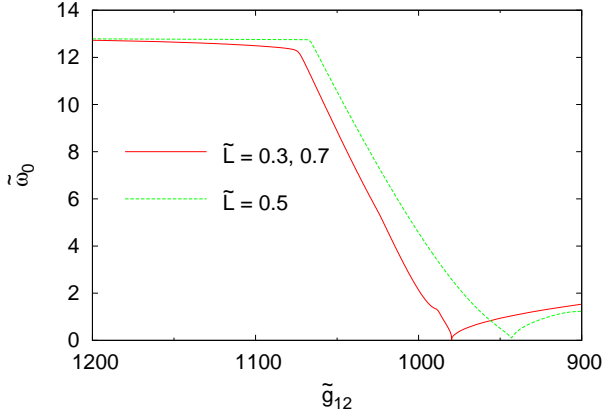}
\caption{
(color online) 
The lowest excitation frequency $\tilde\omega_0$ for the stationary
 state with  fixed $\tilde L$ as a function of $\tilde g_{12}$.
For $\tilde L = 0.3$ and $\tilde L = 0.7$ (solid curve), the excitation
 frequencies are exactly the same.
The parameters are $\tilde g_{11} = \tilde g_{22} = 1000$, $n_1 = 0.9$,
 and $n_2 = 0.1$.
After the transition to the uniform state in the solid curve,
 $\tilde\Omega$ is fixed.
}
\label{f:bogo}
\end{figure}
Figure~\ref{f:bogo} shows the lowest excitation frequency
$\tilde\omega_0$ for $\tilde L = 0.3$, $0.5$, and $0.7$.
The excitation frequencies for $\tilde L = 0.3$ and $0.7$ are exactly
the same, since both states are connected to each other by
Eqs.~(\ref{trans}) and (\ref{Ltrans}).
We find that $\tilde\omega_0$ decreases with $\tilde g_{12}$ and reaches
$\tilde\omega_0 = 0$, at which the adiabatic condition breaks down.
For $\tilde L = 0.3$ and $0.7$ (solid curve), this point ($\tilde g_{12}
\simeq 980$) corresponds to the transition to the uniform state, which
indicates that the localized state cannot transform to the uniform state
adiabatically.
For $\tilde L = 0.5$ (dashed curve), $\tilde\omega_0 = 0$ at $\tilde
g_{12} \simeq 940$.

For example, let us consider a $^{87}{\rm Rb}$ BEC with density
$10^{14}$ ${\rm cm}^{-3}$ in a ring with radius $10$ $\mu {\rm m}$.
In this case the unit of time is $u_t = m R^2 / \hbar \simeq 0.14$ s.
If the typical time scale of change in $\tilde g_{12}$ is 1 s $\sim 10
u_t$, the energy gap $\tilde\omega_0$ must be $\gtrsim 1$ for the
adiabatic condition, and hence $\tilde g_{12}$ can be decreased to
$\simeq 990$ ($\simeq 960$) for the solid (dashed) curve in
Fig.~\ref{f:bogo}.

\subsection{Analysis for rotation frequency}
\label{s:1Danalytic}

In Sec.~\ref{s:1Dground}, we showed that the rotation frequency $\Omega$
changes as the intercomponent interaction $g_{12}$ is changed.
In this subsection, we derive an analytic expression of the rotation
frequency $\Omega$.

We write the wave function as
\begin{equation}
\tilde\psi_j(\theta) = \sqrt{\rho_j(\theta)} e^{i \phi_j(\theta)},
\end{equation}
where $\rho_j(\theta)$ and $\phi_j(\theta)$ are real functions.
We assume that $\rho_2(\theta)$ is localized near $\theta = 0$ and
define $\theta_0$ at which $\rho_2$ decays to zero $\rho_2(\pm \theta_0)
\simeq 0$.
We also assume that the total density is almost constant,
$\rho_1(\theta) + \rho_2(\theta) \simeq (2 \pi)^{-1} \equiv \rho_0$,
and hence $\rho_1(\theta)$ falls only around $\theta = 0$ and
$\rho_1(\theta) \simeq \rho_0$ for $|\theta| > \theta_0$ ($-\pi \leq
\theta \leq \pi$).
We write the phases as
\begin{subequations} \label{phi}
\begin{eqnarray}
\label{phi1}
\phi_1(\theta) & = & \tilde \Omega_0 \theta + f(\theta), \\
\label{phi2}
\phi_2(\theta) & = & \tilde \Omega \theta,
\end{eqnarray}
\end{subequations}
where $\Omega_0$ is a constant determined later.
The function $f(\theta)$ changes only around $\theta = 0$ and satisfies
$f'(\theta) \simeq 0$ for $|\theta| > \theta_0$.
From the single-valuedness of the wave function, Eq.~(\ref{phi1}) gives
\begin{equation} \label{quant}
2 \pi \tilde \Omega_0 + \int_{-\pi}^\pi f'(\theta) d\theta = 2 \ell \pi,
\end{equation}
where $\ell$ is an integer.
Since $\rho_2(\theta)$ vanishes for $|\theta| > \theta_0$, there is no
constraint on Eq.~(\ref{phi2}) for $|\theta| > \theta_0$.
Using $\partial \rho_1 / \partial t = 0$ and Eq.~(\ref{nGP}), we find
\begin{equation} \label{const}
J_1(\theta) - \tilde \Omega \rho_1(\theta) = {\rm const.},
\end{equation}
where $J_j$ is the flux given by
\begin{equation} \label{flux}
J_j(\theta) = \rho_j(\theta) \phi_j'(\theta).
\end{equation}
From Eqs.~(\ref{phi2}), (\ref{const}), and (\ref{flux}), we obtain
\begin{equation} \label{fprime}
f'(\theta) = -(\tilde\Omega - \tilde\Omega_0) \left[
\frac{\rho_0}{\rho_1(\theta)} - 1 \right].
\end{equation}
The angular momentum of each component in Eq.~(\ref{L}) is rewritten as
\begin{subequations} \label{L1L2}
\begin{eqnarray}
\tilde L_1 & = & \int_{-\pi}^\pi \rho_1(\theta) \phi_1'(\theta) d\theta
= \int_{-\pi}^\pi \rho_1(\theta) \left[ \tilde \Omega_0 + f'(\theta)
\right] \nonumber \\
& = & \tilde \Omega_0 - n_2 \tilde \Omega, \\
\tilde L_2 & = & n_2 \tilde \Omega,
\label{L2}
\end{eqnarray}
\end{subequations}
where we used Eqs.~(\ref{phi}) and (\ref{fprime}).
Equation~(\ref{L2}) is consistent with the agreement between the solid
curves and circles in Figs~\ref{f:1D1}-\ref{f:1D3}.
Equation~(\ref{L1L2}) gives
\begin{equation} \label{omega0}
\tilde L = \tilde L_1 + \tilde L_2 = \tilde\Omega_0.
\end{equation}
Using Eqs.~(\ref{quant}), (\ref{fprime}), and (\ref{omega0}), we obtain
\begin{equation} \label{omega}
\tilde\Omega = \ell + (\tilde L - \ell) \left\{ 1 +
\frac{2\pi}{\int_{-\pi}^\pi \left[ \frac{\rho_0}{\rho_1(\theta)} -
1\right] d\theta} \right\}.
\end{equation}

We assume Gaussian density distributions as
\begin{subequations} \label{Gauss}
\begin{eqnarray}
\rho_1(\theta) & = & \rho_0 - \rho_2^{\rm peak} e^{-\theta^2 /
 \sigma^2}, \\
\rho_2(\theta) & = & \rho_2^{\rm peak} e^{-\theta^2 / \sigma^2},
\end{eqnarray}
\end{subequations}
where $\rho_2^{\rm peak} = n_2 / (\sqrt{\pi} \sigma)$ is the peak
density of component 2.
The Gaussian is assumed to be very narrow, $\sigma \ll 1$, so that the
wave functions smoothly connect at $\theta = \pm\pi$.
Substituting Eq.~(\ref{Gauss}) into Eq.~(\ref{omega}), we obtain
\begin{equation} \label{ar}
\tilde\Omega = \ell + (\tilde L - \ell) \left[ 1 +
\frac{1}{n_2 \frac{\rho_0}{\rho_2^{\rm peak}} g_{1/2}
\left( \frac{\rho_2^{\rm peak}}{\rho_0} \right)} \right],
\end{equation}
where the function $g_{1/2}$ is defined by
\begin{equation}
g_{1/2}(z) = \sum_{k = 1}^{\infty} \frac{z^k}{k^{1/2}}.
\end{equation}

\begin{figure}[t]
\includegraphics[width=8.5cm]{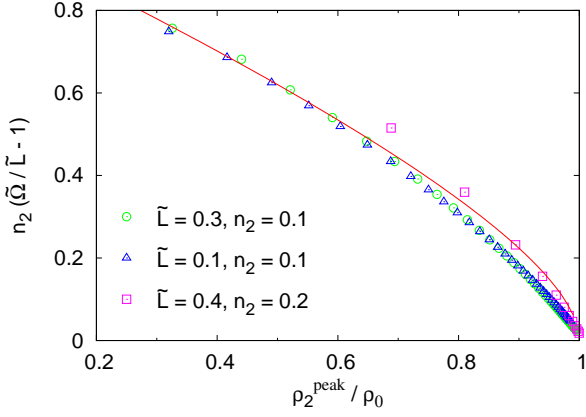}
\caption{
(color online) Relation between the peak density of component 2 and the
 rotation frequency for $(\tilde L, n_2) = (0.3, 0.1)$ (circles), $(0.1,
 0.1)$ (triangles), and $(0.4, 0.2)$ (squares).
The solid curve shows Eq.~(\ref{ar}).
}
\label{f:analytic}
\end{figure}
Figure~\ref{f:analytic} shows $n_2 (\tilde\Omega / \tilde L - 1)$ as a
function of $\rho_2^{\rm peak} / \rho_0 = 2 \pi \rho_2^{\rm peak}$.
The plots show the numerical results obtained by solving Eq.~(\ref{nGP})
and the solid curve shows Eq.~(\ref{ar}) with $\ell = 0$, i.e.,
\begin{equation} \label{univ}
n_2 \left( \frac{\tilde\Omega}{\tilde L} - 1 \right) =
 \frac{\frac{\rho_2^{\rm peak}}{\rho_0}}{g_{1/2}\left(\frac{\rho_2^{\rm
peak}}{\rho_0} \right)}. 
\end{equation}
We find that the plots with different parameters fit the universal curve
given by Eq.~(\ref{univ}).
According to Eqs.~(\ref{trans}) and (\ref{Ltrans}), numerical results
with $\tilde L \rightarrow \ell - \tilde L$ also agree well with
Eq.~(\ref{ar}).

\section{Three dimensional toroidal trap}
\label{s:3D}

\begin{figure}[t]
\includegraphics[width=7.5cm]{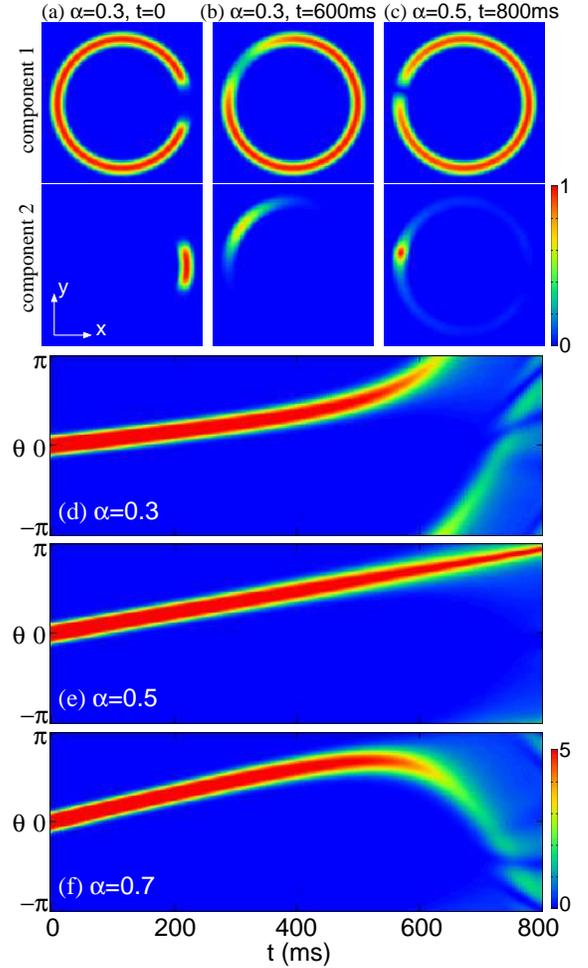}
\caption{
(color online) Column density profiles $\int |\psi_j|^2 dz$ for (a)
 $(\alpha, t) = (0.3, 0)$, (b) (0.3, 60 ms), and (c) (0.5, 80 ms).
The field of view is $25 \times 25$ $\mu{\rm m}$.
Density profiles of component 2 at $r_\perp = R$ and $z = 0$ for (d)
 $\alpha = 0.3$, (e) $\alpha = 0.5$, and (f) $\alpha = 0.7$.
The density is normalized by $N / R^2$ in (a)-(c) and by $N / R^3$ in
(d)-(f).
}
\label{f:3D}
\end{figure}
We perform a numerical calculation of full 3D real-time propagation in a
realistic situation.
We consider a BEC of $^{87}{\rm Rb}$ atoms in the hyperfine states $|F,
m_F \rangle = |2, 2 \rangle$ and $|2, 1 \rangle$, which we refer to as
components 1 and 2, respectively.
The scattering lengths between component 1 is $a_{11} = a_4$ and that
between component 2 is $a_{22} = (3 a_2 + 4 a_4) / 7$, where $a_S$ is
the $s$-wave scattering length between two atoms with total spin $S$.
Their difference is $a_{11} - a_{22} = 3 (a_4 - a_2) / 7 \simeq 2.98
a_{\rm B}$ with $a_{\rm B}$ being the Bohr radius, where we used the
value measured in Ref.~\cite{Widera}.
We therefore use the scattering lengths $a_{11} = 100 a_{\rm B}$ and
$a_{22} = (100 - 2.98) a_{\rm B}$.
The intercomponent scattering length $a_{12}$ is assumed to be variable.
The number of atoms is $N = N_1 + N_2 = 2 \times 10^5$ with $N_1 / N_2 =
9$.
We employ a toroidal-shaped trap as
\begin{equation}
V_1 = V_2 = \frac{1}{2} m \omega^2 \left[ \left( r_\perp - R \right)^2 +
z^2 \right],
\end{equation}
with a frequency $\omega = 2\pi \times 1$ kHz and a radius $R = 10$
$\mu{\rm m}$, where $r_\perp = (x^2 + y^2)^{1/2}$.

We first prepare the nonrotating ground state $\Psi_j(\bm{r})$ for
$a_{12} = 100 a_{\rm B}$ by the imaginary-time propagation of
Eq.~(\ref{GP}), which is phase separated as shown in Fig.~\ref{f:3D}
(a).
We then imprint the phase as $\Psi_j(\bm{r}) e^{i \alpha \theta}$ with
$0 < \theta < 2\pi$ for $j = 1$ and $-\pi < \theta < \pi$ for $j = 2$,
where a real parameter $\alpha$ determines the initial angular momentum.
In an experiment, such phase imprinting is done by, e.g., spatially
modulated laser fields~\cite{Matthews}.
If $\alpha$ is not an integer, the phase of component 1 (2) jumps at
$\theta = 0$ ($\theta = \pi$), at which the density vanishes.
Using this initial state, we study the dynamics of the system by solving
Eq.~(\ref{GP}) in a laboratory frame ($\Omega = 0$) with the
pseudospectral method.
The intercomponent scattering length is decreased as
\begin{equation}
a_{12}(t) = 100 (1 - 0.05 t / t_{\rm d}) a_{\rm B}
\end{equation}
with $t_{\rm d} = 1$ s.

Figures~\ref{f:3D} (d)-\ref{f:3D} (f) show the dynamics of the density
profile of component 2 on the circumference of the torus,
$|\psi_2(r_\perp = R, z = 0, t)|^2$.
For $t \lesssim 300$ ms, the rotation frequencies are almost constant.
At a later time, the rotation of the system accelerates for $\alpha =
0.3$ [Fig.~\ref{f:3D} (d)], and reverses for $\alpha = 0.7$
[Fig.~\ref{f:3D} (f)].
For $\alpha = 0.5$, the solitonic structure remains until $t = 800$ ms
as shown in Fig.~\ref{f:3D} (c).
These behaviors are consistent with the 1D results in
Figs.~\ref{f:1D1}-\ref{f:1D3}.
Thus, we have shown that the rotation properties studied for a 1D ring
in Sec.~\ref{s:1D} can be observed in a 3D toroidal geometry
experimentally.

\section{Conclusions}
\label{s:conc}

We have studied a two-component BEC rotating in a toroidal trap, in
which the intercomponent interaction is controlled.
When the two components are phase separated along the circumference of
the torus, the system can rotate without topological constraint.
As the intercomponent repulsion is decreased and the two components
mix, the topological constraint is imposed on the phase of the wave
function.
Thus, the interplay between the quantization of circulation and the
angular momentum conservation exhibits nontrivial phenomena.

We found that as the two components become miscible, the system goes to
the states with circulations $\Gamma_j$ depending on the initial angular
momentum $L$.
For a small initial angular momentum (e.g., $L = 0.3 N \hbar$), the
major component 1 goes to $\Gamma_1 = 0$ while the minor component 2
goes to $\Gamma_2 > 0$, resulting in the acceleration of rotation
(Fig.~\ref{f:1D1}).
For a larger angular momentum ($L = 0.7 N \hbar$), the circulations go
to $\Gamma_1 = 1$ and $\Gamma_2 < 0$, and the system counterrotates
(Fig.~\ref{f:1D3}).
For $L = 0.5 N \hbar$, the stable dark soliton is generated
(Fig.~\ref{f:1D2}).
The Bogoliubov analysis gives the adiabatic condition for the change in
the intercomponent repulsion (Fig.~\ref{f:bogo}), and the Gaussian
analysis gives the expression of the rotation frequency (Eq.~(\ref{ar})
and Fig.~\ref{f:analytic}).
The full 3D numerical analysis has shown that the predicted phenomena
can be observed in a realistic situation in an experiment.

\begin{acknowledgments}  
This work was supported by Grants-in-Aid for Scientific Research (No.\
 20540388 and No.\ 22340116) from MEXT and Highly Talented Young
 Researcher under MEXT's Program ``Special Coordination Funds for
 Promoting Science and Technology.''
\end{acknowledgments}

\end{document}